\documentclass{ws-procs9x6}

\begin{document}

\title{On the ion-mediated interaction between protein and DNA}

\author{M. Barbi}

\address{CNRS LPTMC UMR 7600, Universit\'e Pierre et Marie Curie-Paris 6\\ 4
place Jussieu, 75252 Paris Cedex 05, France\\
E-mail: barbi@lptmc.jussieu.fr}

\author{F.Paillusson}

\address{Department of Chemistry, University of Cambridge, Lensfield Road\\
CB2 1EW, Cambridge, UK\\
E-mail: fp286@cam.ac.uk}

\begin{abstract}
The mechanism allowing a protein to search of a target sequence on DNA is currently described as an intermittent process composed of 3D diffusion in bulk and 1D diffusion along the DNA molecule. Due to the relevant charge of protein and DNA,
electrostatic interaction should play a crucial role during this search. 
In this paper, we explicitly derive the mean field theory allowing for a description of the protein-DNA electrostatics in solution. This approach leads to an unified model of the search process, where 1D and 3D diffusion appear as a natural consequence of the diffusion on an extended interaction energy profile.
\end{abstract}

\keywords{DNA; Proteins; ionic liquids; modelling}

\bodymatter

\section{Introduction}\label{barbi:sect01}
Many proteins in living cells have to search specific, short sequences on long DNA molecules in order to perform their biological task. Such {\it DNA-binding proteins} have proven to be very efficient in searching their target: their association constants can be two orders of magnitude higher than what is expected from a simple 3D diffusion\cite{Rig70,Richter}. It has been suggested\cite{von89,ben04} that such a rapid reaction rate can results from an {\it intermittent diffusion}, swapping between a 1D diffusion along DNA -- or {\it sliding} -- and a 3D diffusion in solution -- or {\it jumping}. An increasing number of single particle experiments has been able to evidence sliding, confirming this scenario\cite{Shi99,Bla06,Elf07,Deb08}. 
Experiments also show that both the sliding and the jumping result to be sensitive to the salt concentration\cite{Rig70,Bla06,Deb08}. This  supports the idea that electrostatics is involved to some extent in the intermittent behaviour, with a probable role for the solution ions.
 
In this paper, we shall first recall the statistical mechanics of ions in solution for a given fixed charge distribution. We will then introduce a toy model for a DNA-protein system\cite{PRL,Paillusson11} for which we will discuss some unexpected features. This will finally allow us to gain more insights about the physics at play during the search of target by a protein  and to propose a method to get further insights on protein physical properties.

\section{Statistical mechanics of electrolytes}\label{barbi:sect02}
In this section we consider two fixed macro-ions confined in a domain $\Sigma \subset \mathbb{R}^3$ that contains an electrolyte solution. We denote $\rho_f$ the  charge density carried by these macro-ions. Within the framework of statistical mechanics \cite{McQuarrie}, the position $\mathbf{R}_j$ of an ion $j$ is a random 
vector that can take any value $\mathbf{r}_i$ belonging to a subset $\Omega$ of $\Sigma$. The set of $N$ 
particle positions $\{\mathbf{r}_1,..,\mathbf{r}_N \}$ gotten at each {\it trial} for $N$ ions in the system will be called an {\it ionic configuration} and denoted $\mathcal{C}$. In thermodynamic equilibrium, the conditional probability density to get a specific configuration $\mathcal{C}$ knowing that the system is at temperature $\beta^{-1}/k_B$ and confined in the domain $\Omega$ of volume $V$ reads:
\begin{equation}
 p(\mathcal{C}|\beta, \mathbf{N} ,V) \equiv \mathbf{1}_{\mathcal{C} \in \Omega^N}\: e^{-\beta H(\mathcal{C})}/Q_{[\beta, \mathbf{N} ,V]} \label{barbi01}
\end{equation}where $\mathbf{1}_{\mathcal{C} \in \Omega^N}$ is the characteristic function that is zero if any of the ions is outside $\Omega$ and one otherwise. In Eq.~\eqref{barbi01}, we introduced the normalization functional:
\begin{equation}
 Q_{[\beta, \mathbf{N}, V]} \equiv \frac{1}{\prod_{\alpha}^m N_{\alpha}! \Lambda_{\alpha}^{3N_{\alpha}}}\int_{\Omega^N} \:d^{3N}(\mathcal{C})\:e^{-\beta H(\mathcal{C})} \label{barbi02}
\end{equation}with $d^{3N}(\mathcal{C})$ being the 3N-dimensional Lebesgue measure on $\mathbb{R}^{3N}$, $m$ is the number of different ionic species in solution, $\Lambda_{\alpha}$ is the de Brooglie wavelength of the species $\alpha$ and the sets $\mathbf{N}=\{N_{\alpha}\}_{\alpha=1..m} \:\in\:\mathbb{N}^m$ and $\mathbf{q}=\{q_{\alpha}\}_{\alpha=1..m}\:\in\:\mathbb{R}^m$ characterize the {\it ionic composition} of the mixture.
 The real valued function $H(\mathcal{C})$ in Eqs.~\eqref{barbi01} and~\eqref{barbi02} is the energy of the system for a given configuration $\mathcal{C}$ that defines the model used. Here, we rely on the so called {\it Restricted Primitive Model} (RPM)\cite{Caillol04}, defined as:
 \begin{equation}
  H(\mathcal{C}) = \frac{1}{2}\int_{\Sigma} \:d^6rr' \frac{\rho_{\mathcal{C}}(\mathbf{r})\rho_{\mathcal{C}}(\mathbf{r'})}{4 \pi \varepsilon |\mathbf{r}-\mathbf{r'}|} + \sum_{j<k}v_{HS}(j,k) \label{barbi03}
 \end{equation}
 where 
 $\varepsilon$ is the dielectric permittivity of water and where we use the Dirac delta ``function'' to define the charge density $\rho_{\mathcal{C}}(\mathbf{r}) \equiv \sum_j \:e\:q_j \delta(\mathbf{r}-\mathbf{r}_j) + \rho_f(\mathbf{r})$. The second term in Eq.~\eqref{barbi02} stands for a hard-sphere repulsion such that $e^{-\beta v_{HS}(j,k)}$ behaves as $\Theta(|\mathbf{r_j}-\mathbf{r_k}|- D)$ where $\Theta(x)$ is the Heavyside step function and $D$ the diameter of the ions\footnote{The diameter $D$ is such that we can assume each ionic species to be in a stable gas phase in solution.}. From Eq.~\eqref{barbi02}, one can then perform a {\it Hubbard-Stratonovich} transform that effectively breaks Coulomb pair interactions into a one-body potential $\phi$ \cite{Martin99,Caillol04}  \footnote{The measure $\mathcal{D}[\phi]$ can be thought of as the limit of the measure $\prod_{k}^{(L/\epsilon+1)^3} d\gamma(\phi_k)$ --- $\gamma$ being a complex measure --- characterizing field configurations on a 3D-lattice of size $L$ and lattice spacing $\epsilon$ when the latter tends to zero.}:
\begin{equation}
 e^{-\frac{\beta}{2}\int_{\Sigma} \:d^6rr' \frac{\rho_{\mathcal{C}}(\mathbf{r})\rho_{\mathcal{C}}(\mathbf{r'})}{4 \pi \varepsilon |\mathbf{r}-\mathbf{r'}|}} \equiv  \int \frac{\mathcal{D}[\phi]}{\mathcal{Z}[0]}e^{-\frac{\beta}{2}\int_{\Sigma} d^6rr'\:\phi(\mathbf{r})G^{-1}_{\mathbf{r},\mathbf{r'}}\phi(\mathbf{r'})-i\beta\int d^3r\phi(\mathbf{r})\rho_{\mathcal{C}}(\mathbf{r})} \label{barbiHS}
\end{equation}where $i^2=-1$ and $G^{-1}_{\mathbf{r},\mathbf{r'}}= -\varepsilon \Delta \delta(\mathbf{r}-\mathbf{r'})$ \cite{Martin99}. 
The factor $\mathcal{Z}[0]$ is the normalization factor for the free field $\phi$ in absence of $\rho_{\mathcal{C}}$. 
If we introduce the Gaussian average over configurations of the field $\phi$, $\langle . \rangle_{\phi}$,
then the r.h.s. of Eq.~\eqref{barbiHS} reads $\langle e^{-i\beta \int d^3r\:\rho_{\mathcal{C}}\phi}\rangle_{\phi}$. It is 
now
convenient to swap to a grand canonical ensemble where the composition $\mathbf{N}$ is a random vector taking any value in $\mathbb{N}^m$ with a probability weight $e^{\beta\mathcal{M}\cdot \mathbf{N}} \equiv e^{\beta \sum_{\alpha}\mu_{\alpha}N_{\alpha}}$ set by $m$ {\it chemical potentials} $\mathcal{M}=\{\mu_{\alpha} \}_{\alpha=1..m}\:\in \:\mathbb{R}^m$, each of which corresponds to a particular ionic species. In this case, the normalization functional writes:
\begin{equation}
 Q_{[\beta, \mathcal{M}, V]} \equiv \sum_{\mathbf{N}} e^{\beta \mathcal{M}\cdot \mathbf{N}}Q_{[\beta, \mathbf{N}, V]} \label{barbi05}
\end{equation}
Inserting the partition function~\eqref{barbi02} (after performing~\eqref{barbiHS}) into Eq.~\eqref{barbi05} yields:
\begin{equation}
 Q_{[\beta, \mathcal{M}, V]} = \left< Q_{[\beta,\mathcal{M}(i\phi),V]}^{HS}e^{-i\beta \int d^3r\:\rho_f \phi}\right>_{\phi} \label{barbi06bis}
\end{equation}
where $Q_{[\beta,\mathcal{M}(i\phi),V]}^{HS}$ is the grand partition function \cite{McQuarrie} of a mixture of {\it bare} hard spheres with the set of chemical potentials $\mathcal{M}(i\phi)=\mathcal{M}-i e\mathbf{q}\phi$. We now use the fact that the mixture is dilute by approximating $Q_{[\beta,\mathcal{M}(i\phi),V]}^{HS}$ by the first term of its Mayer expansion \cite{McQuarrie} i.e. $Q_{[\beta,\mathcal{M}(i\phi),V]}^{HS} \approx \exp(\int d^3 r\:\sum_{\alpha} e^{\beta(\mu_{\alpha}-ieq_{\alpha}\phi)}/\Lambda_{\alpha}^3)$. Eq.~\eqref{barbi06bis} becomes then explicitely:
\begin{equation}
 Q_{[\beta, \mathcal{M}, V]} = \int \frac{\mathcal{D}[\phi]}{\mathcal{Z}[0]} e^{-\beta \int_{\Sigma} d^3r\:[\varepsilon \frac{(\nabla\phi)^2}{2} - \beta^{-1} \mathbf{1}_{\mathbf{r}\in\Omega}\sum_{\alpha = 1}^m e^{\beta (\mu_{\alpha}-i e q_{\alpha} \phi)}/\Lambda_{\alpha}^3 + i\rho_f \phi] } \label{barbi06}
\end{equation}
Note that the quadratic term in $\phi$ from Eq.~\eqref{barbiHS} has now become quadratic in $\nabla\phi$, by applying twice the divergence theorem\footnote{The boundary terms arising from this theorem do not contribute since global electro-neutrality is assumed in $\Sigma$.}.

The so called {\it Poisson-Boltzmann} (PB) theory can now be readily gotten from Eq.~\eqref{barbi06} by formally using a functional saddle point or {\it mean field} approximation \cite{Orland99,Caillol04}. The grand potential defined as $\mathcal{G}_{[\beta, \mathcal{M}, V]} \equiv -\beta^{-1} \ln Q_{[\beta, \mathcal{M}, V]} $ reads then:
\begin{equation}
 \mathcal{G}_{[\beta, \mathcal{M}, V]} \stackrel{_{MF}}{=}  - \int_{\Sigma} d^3r\:\big[\varepsilon \frac{(\nabla\varphi)^2}{2} + \beta^{-1}\mathbf{1}_{\mathbf{r}\in\Omega}\sum_{\alpha = 1}^m \frac{e^{\beta (\mu_{\alpha}-h- e q_{\alpha} \varphi)}}{\Lambda_{\alpha}^3} -\rho_f \varphi \big]  \label{barbi07}
\end{equation}
where the sign $\stackrel{_{MF}}{=} $ stands for an equality within the mean field approximation and where $\varphi \equiv i \phi_{s}$, $\phi_s$ being the field evaluated at the saddle point.  With the new field $h$ introduced in~\eqref{barbi07} and $\rho_f$,  $\mathcal{G}_{[\beta, \mathcal{M}, V]}$ can be seen as a generating functional from which one can get grand canonical averages of meaningful quantities. In particular $\langle i\phi\rangle_{\beta,\mathcal{M},V} \equiv -(\delta \mathcal{G}_{\beta, \mathcal{M}, V}/\delta \rho_f)_{h=0}\stackrel{_{MF}}{=} \varphi$ and $\langle \rho_{\mathcal{C}}^{\alpha} \rangle_{\beta, \mathcal{M}, V} \equiv -e q_{\alpha}(\delta \mathcal{G}_{\beta, \mathcal{M}, V}/\delta[\beta (\mu-h)])_{h=0} \stackrel{_{MF}}{=} eq_{\alpha}\mathbf{1}_{\mathbf{r}\in\Omega} e^{\beta \mu_{\alpha}}e^{-\beta e q_{\alpha}\varphi}/\Lambda_{\alpha}^3$. 
The average charge density $\langle \rho_{\mathcal{C}}^{\alpha} \rangle_{\beta, \mathcal{M}, V}$ has to be a real number which implies then that $\varphi$ has to be a real field. It is therefore common to use Eq.~\eqref{barbi07} as a functional of the real valued function $\varphi$ called the {\it Poisson-Boltzmann functional} that has to be extremalized numerically to find the most probable electrostatic field $\varphi$ and its corresponding charge density \cite{Maggs12}. An equivalent way to look at it is to realize that if $\varphi$ maximizes the PB functional:  the functional derivative of the latter with respect to the former has therefore to be zero. This gives rise to an Euler-Lagrange type of equation called the {\it Poisson-Boltzmann equation} to be solved for $\varphi$:
\begin{equation}
 \Delta \varphi = -\frac{1}{\varepsilon}\left(\mathbf{1}_{\mathbf{r}\in\Omega}\sum_{\alpha=1}^m e q_{\alpha}\frac{e^{\beta \mu_{\alpha}}}{\Lambda_{\alpha}^3}e^{-\beta e q_{\alpha}\varphi} + \rho_f\right) \label{barbi10}
\end{equation}which turns out to be the most common route used to determine the average electrostatic potential $\varphi$.

\section{Modelling protein-DNA interactions}\label{barbi:sect03}
\begin{figure}[htbp]
\begin{center}
\textbf{(a)} \includegraphics[width=.3\textwidth,keepaspectratio=true]{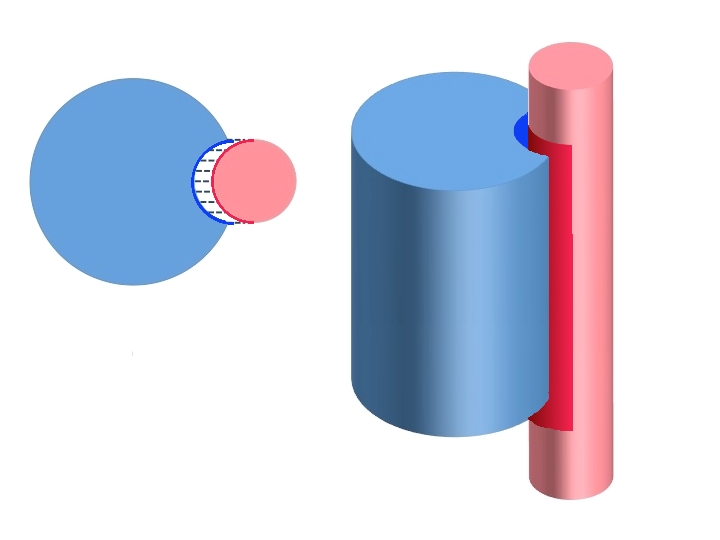}
\textbf{(b)} \includegraphics[width=.3\textwidth,keepaspectratio=true]{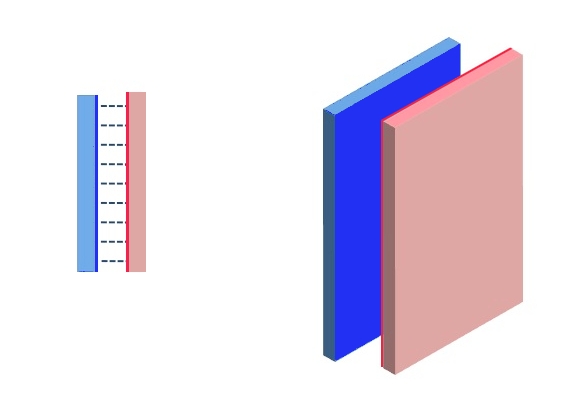}\vspace{4mm}

\textbf{(c)}\includegraphics[width=.8\textwidth,keepaspectratio=true]{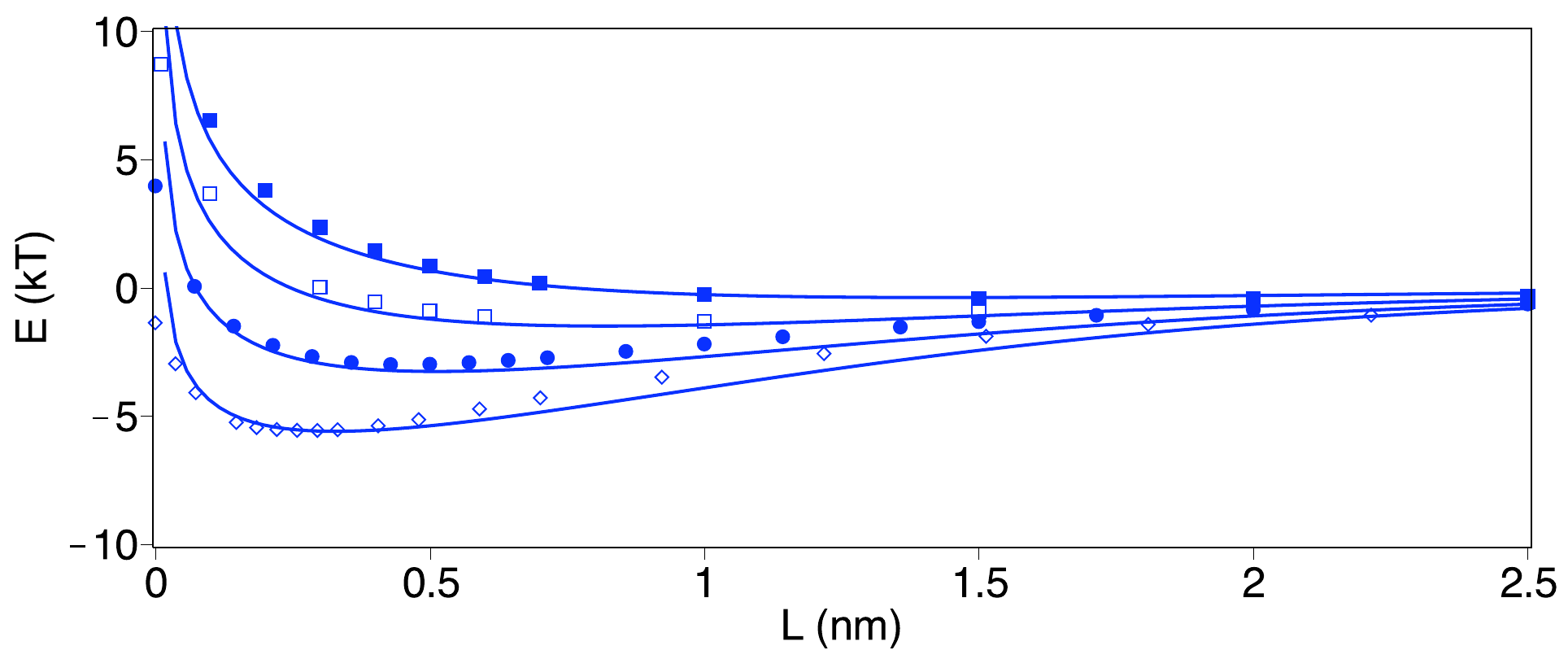}
\end{center}
\caption{\textbf{(a)} Toy model for a protein interacting with a DNA segment in solution.
\textbf{(b)} The toy model can be mapped onto two oppositely but non-symmetrically charged plates in solution.
\textbf{(c)} Free energy profiles obtained by integrating the PB equation for the plate plate system with (physiological) monovalent salt concentration of $ 0.1\:\rm mol/L$, and for 4 different protein charge densities in the range from 0.06 to 0.3 times the absolute value of the DNA charge density. 
}
\label{fig:barbi01}
\end{figure}

Many DNA binding proteins happen to have a concave shape matching that of DNA. This is believed to optimize the recognition at the target site. In addition, these proteins need to be positively charged otherwise they would be repelled by DNA's high negative charge. In a previous work, we have suggested a toy model to study the relevance of the geometry in DNA-protein non specific interactions~\cite{PRL,Paillusson11}. In this model, depicted in Fig. \ref{fig:barbi01} (a), the model DNA (MDNA) is a uniformly charged cylinder and the model protein (MP) is a cylinder of larger radius but with an indentation of cylindrical shape that matches exactly the DNA shape, and is positively charged at the interface. They are also immersed in a RPM  of a symmetric 1:1 electrolyte as described in Eq.~\eqref{barbi03}. In the terminology of section~\eqref{barbi:sect02}, the charge density on these macromolecules -- for a given distance $\mathcal{L}$ between them -- corresponds to $\rho_f(\mathcal{L})$, the whole system is in a domain $\Sigma$ and $\Omega(\mathcal{L})$ is the accessible region to ions i.e. anywhere in $\Sigma$ except inside the macromolecules. 
The $+$ and $-$ ion bulk concentrations are assumed to be the same and  denoted $n_b \equiv e^{\beta \mu_{\pm}}/\Lambda_{\pm}^3$.
The grand potential $ \mathcal{G}_{[\beta, \mathcal{M}, V]}$ of this whole system is implicitly dependent on the domain $\Omega(\mathcal{L})$ and the distribution $\rho_f(\mathcal{L})$. In fact, if we were to consider $\mathcal{L}$ as a random variable subject to thermal fluctuations, the grand potential $\mathcal{G}_{[\beta, \mathcal{M}, V]}(\mathcal{L})$ would act as an {\it effective interaction energy} \cite{Voth} between the MDNA and the MP such that each value $l$ of $\mathcal{L}$ appears with a probability weight $e^{-\beta \mathcal{G}_{[\beta, \mathcal{M}, V]}(l) }$.

Monte Carlo (MC) simulations of the system described above have been performed to compute exactly $\mathcal{G}_{[\beta, \mathcal{M}, V]}(\mathcal{L})$ via a so called {\it thermodynamic integration} \cite{FrenkelSmit}. It was found~\cite{PRL,Paillusson11} that $\mathcal{G}_{[\beta, \mathcal{M}, V]}(l)$ is an increasing function if $l \:\in\:[l^*,+\infty[$ and a decreasing function if $l\:\in\:[0,l^*[$. In physical terms, $l^*$ corresponds to a stable equilibrium distance between the MP and the MDNA segment~\cite{Pai09}. It was also shown that the profile $\mathcal{G}_{[\beta, \mathcal{M}, V]}(l)$ gotten from MC simulations could be matched with a PB theory for two plates (as shown in Fig.~\ref{fig:barbi01} (b)) by solving the PB equation~\eqref{barbi10} for $\varphi$ and evaluating the expression~\eqref{barbi07} for every $l$. This ``mapping'' between the MC implementation of the toy model of Fig.~\ref{fig:barbi01} (a) and a PB treatment of a two plate system is valid provided effective charge densities -- related to that of the MDNA and the MP -- are used for the plates~\cite{PRL,Paillusson11}. The actual values of these parameters depend on the particular modelling of $\rho_f$ used in the MC simulations and therefore do not provide at the moment any more insights about what is happening in the system.

Overall, the intermittent behaviour observed for the DNA-protein system can be rationalized by considering the random nature of $\mathcal{L}$ and treating properly the physics of the ions: 
in a non-specific DNA-protein bound state, sliding is possible at the equilibrium distance $l^*$, while thermal fluctuations can still make the protein escape from DNA, in which case it would perform a jump.
Let us finally note that, in practice, the density $\rho_f$ cannot be measured experimentally, and is often inferred from structural data. Interestingly, the 
simple planar PB description  that we have introduced is not only able to capture this physics, but also provides analytical expressions -- as a function of effective charge densities $\rho_f$  -- for both $l^*$ and $\mathcal{G}_{[\beta, \mathcal{M}, V]}(l^*)$.
Since these quantities directly determine the kinetic behaviour of the protein, 
a comparison with independent structural and dynamical data from experiments may be used as a simple alternative to estimate coarse grained surface densities at the protein-DNA interface.

\bibliographystyle{ws-procs9x6}
\bibliography{contrib_barbi}

\end{document}